\documentclass[twocolumn]{article}

\usepackage[normalem]{ulem}
\usepackage{parskip}
\usepackage{pdfpages}
\usepackage{url}

\begin{document}

\title{Cyclone: High Availability for Persistent Key Value Stores}
\author{Amitabha Roy \\ amitabha.roy@gmail.com 
	     \and Subramanya R. Dulloor \\ dulloor@gmail.com}

\date{}
\maketitle

\begin{abstract}
\vspace{0.05in}
Persistent key value stores are an important component of many distributed data
serving solutions with innovations targeted at taking advantage of growing flash
speeds. Unfortunately their performance is hampered by the need to maintain and
replicate a write ahead log to guarantee availability in the face of machine and
storage failures. Cyclone is a replicated log plug-in for key value stores that
systematically addresses various sources of this bottleneck. It uses a small
amount of non-volatile memory directly addressable by the CPU - such as in the
form of NVDIMMs or Intel 3DXPoint - to remove block oriented IO devices such as
SSDs from the critical path for appending to the log. This enables it to address
network overheads using an implementation of the RAFT consensus protocol that is
designed around a userspace network stack to relieve the CPU of the burden of
data copies.  Finally, it provides a way to exploit the parallelism available in commodity NICs.  
Cyclone is able to replicate millions of small updates per second using only commodity 10
gigabit ethernet adapters. As a practical application, we use it to improve the
performance (and availability) of RocksDB, a popular persistent key value store
by an order of magnitude when compared to its own write ahead log without
replication.
\end{abstract}

\section{Introduction}
Persistent key value stores are an important component of datacenter scale
storage services. Key value stores such as Rocksdb~\cite{rocksdb},
LevelDB~\cite{leveldb} and FloDB~\cite{flodb} represent significant efforts on
both engineering and research fronts. These key value stores include
sophisticated in-memory data structures built around Log Structured Merge (LSM)
trees~\cite{lsmtree} and are heavily tuned to extract maximum performance from
flash-based solid state drives (SSDs).

\begin{figure}
\centering \includegraphics[scale=0.6]{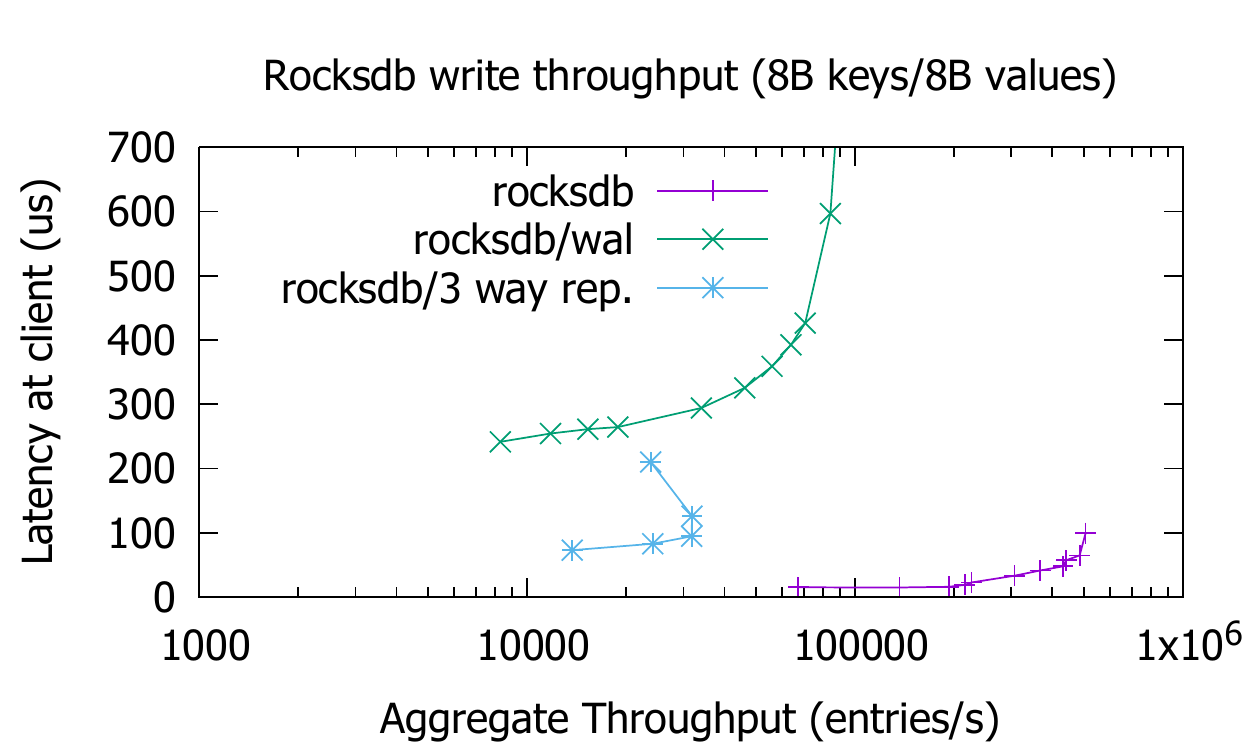}
\caption{Rocksdb write ahead logging impact}
\label{fig:problem}
\end{figure}

These key value stores however tend to ignore an important component: the write
ahead log. A machine or storage failure leading to total loss or temporary
unavailability of data is unacceptable in services where high availability and
revenue are interconnected. Key value stores therefore usually incorporate
support for a write ahead log that if replicated and kept durable for every
appended update provides the necessary high availability. Unfortunately, the
write ahead log is a performance achilles heel for these systems, eclipsing much
of the work on improving the performance of the LSM component. To illustrate the
impact of the write ahead log, consider Figure~\ref{fig:problem}. The line
marked 'Rocksdb' shows the performance of Rocksdb without the write ahead
log. The performance when persisting the write ahead log without replicating it
is shown as the line marked 'rocksdb/WAL'. The line marked 'rocksdb/3 way rep.'
is for simply replicating the log three ways without persisting it, using
RAFT~\cite{raft} running over TCP/IP. Either persisting every update to the log
or replicating it using TCP/IP causes performance to drop by an order of
magnitude (note the log scale on the x-axis). It is therefore no surprise that
deployments of Rocksdb often turn off the write ahead log~\cite{samza},
depending on upper layers to provide availability by co-ordinating replicas. On
the other side of the spectrum, key value store research prototypes such as
FloDB~\cite{flodb} turn off the write ahead log to be able to showcase benefits
of sophisticated extensions to LSM data structures.

These two performance bottlenecks illustrated in Figure~\ref{fig:problem} are
\emph{not independent}, they cannot be addressed in isolation. In particular,
durability semantics mean that the key-value store cannot respond to the client
until the operation has been persisted to the log at a majority of replicas,
since it must be completed even on a crash and subsequent recovery. Therefore,
addressing the network bottleneck is pointless without also addressing the
storage bottleneck. Cyclone shows how a small amount of non-volatile memory can
be used to address both bottlenecks.

Cyclone is a high speed strongly consistent replicated write ahead logging
service specialized for key value stores such as Rocksdb. Cyclone entirely
closes the performance gap due to both the storage and the network shown in
Figure~\ref{fig:problem}. We show that Cyclone achieves performance transparent
replication in that it can provide both a persistent write ahead log and
replication without compromising on RocksDB's performance.

The first key contribution of Cyclone is demonstrating the benefit of a small
amount of non-volatile memory when combined with the flash SSD based persistent
logs.  The primary cause of slowdown when turning on a write ahead log on a
single machine in Figure~\ref{fig:problem} is the need to do synchronous block
IO for every update \emph{regardless of the size of the update}. SSDs provide
the maximum throughput when writing large chunks (usually more than 4KB) of
data. This is due to the fixed cost of a round trip to the SSD through the IO
interface (NVMe in our case), regardless of the volume of data being
synchronously persisted. Small updates (sum of key and value sizes) are
surprisingly common in observed workloads for key value pairs. For example,
Facebook~\cite{fb_workload_analysis} reports that 90\% of the space allocated in
key value stores is for data items under 500 bytes in size.

Cyclone avoids this overhead by making use of a small amount of directly
attached non-volatile memory (NVM), also called persistent memory, on the
server.  This can be in the form of NVDIMMs (DIMMs with an added ultracapacitor
to dump state to a small amount of attached flash) -or- newer and potentially
cheaper (but slower) forms of persistent memory~\cite{pmfs}.
%For current deployments, NVDIMMs are significantly more expensive than standard
%DIMMs, which means that they will %likely be available only in smaller
%quantities compared to flash.
In cases where the log does not entirely fit in NVM, we periodically drain the
NVM log into a log placed on a standard SSD on the system.

The second key contribution of this work is in showing that the throughput
bottlenecks in the network stack can be mitigated entirely in software, once
block IO is removed from the critical path. Existing work on log replication
using state machine replication protocols such as Paxos~\cite{paxos} and
RAFT~\cite{raft} has not - thus far - addressed high performance replication in
the local area network of a datacenter using commodity networks.  Research
focusing on addressing network latencies by reducing the number of
network hops with new protocols such as Fast Paxos~\cite{fast-paxos} or network
switch modifications such as No Paxos~\cite{nopaxos} do not address throughput
bottlenecks due to network stack related inefficiencies in the participating
nodes of the quorum -- a more urgent problem with local area replication. On the
other hand, work done to address network related inefficiencies start with the
assumption that the bottleneck to good throughput on the network is the CPU and
therefore one must either resort to network offload mechanisms such as RDMA~\cite{dare,
  farm, faast} or offload the protocol entirely to an
FPGA~\cite{consensus_box}. We demonstrate in this paper that this is not the
case. Rather, once persistence is provided by directly attached non-volatile
memory, thereby eliminating block IO from the persistence step in replication
protocols, the problem becomes akin to multicasting the same log entry
from the leader to all followers. We design Cyclone's network stack around well
known principles in the software packet switching community. In particular, a
careful implementation of a consensus protocol that relieves the CPU of the
responsibility of data movement permits a high performance implementation
entirely in software that addresses the order of magnitude network related
performance gap in Figure~\ref{fig:problem} using only commodity 10 Gigabit
ethernet.

Both the storage (NVM, SSD) as well as the NIC expose significant parallelism
that is left unused when Cyclone is used to replicate a single log. The third
key contribution of this work is to show how one can exploit this parallelism in
Cyclone automatically creates and manages multiple logs, mapping them to the 
available parallelism in the system via multiple instances of the same consensus 
protocol.

The rest of the paper is organized as follows. We describe Cyclone's system
architecture in Section~\ref{sec:sysarch}. We describe the two level log
structure in Section~\ref{sec:storage}. The set of optimizations to the network
stack is described in Section~\ref{sec:network}. We show how Cyclone can be
scaled to make use of available parallelism on the system in
Section~\ref{sec:parallelism}. A detailed evaluation of Cyclone's replication
performance is provided in the context of RocksDB - a popular persistent key
value store - in Section~\ref{sec:evaluation}. We then discuss related work
before concluding.

\section{System Architecture}
\label{sec:sysarch}
A persistent key value store durably stores key-value mappings accessible
through a simple interface:

\begin{itemize}
\item GET(K): returns value corresponding to key K
\item PUT(K, V): sets the value of key K to value V
\item DELETE(K): deletes the key K
\end{itemize}

Key value stores such as RocksDB also support atomic writes ({\tt PUT}
operations) to multiple keys, called batched writes, and the ability to take
snapshots.

Cyclone integrates with key value stores as both a client and server side
library. Cyclone replicates the key value store across a set of replica
servers. The client side library sends \emph{all} requests - reads and updates -
to a distinguished leader replica.  On the server side Cyclone accepts requests
from clients, calls into the key value store, and returns a response to the
client. Before executing any PUT or DELETE request, Cyclone appends it to a
durable log and replicates the request to the logs of follower replicas. The
request is considered replicated once acknowledged by a majority of
replicas. Follower replicas apply updates from the log in order.

Cyclone uses the RAFT consensus protocol~\cite{raft} to keep the logs in sync
across the replicas. On failure of the leader replica, a new leader is
automatically elected -- RAFT ensures that the new leader has the most up to
date log. The Cyclone client library automatically locates the new leader on a
failover, resulting in only a brief interruption in service for
clients.  Cyclone guarantees that the order of operations to any particular key 
are linearizable. A key sees the exact same sequence of update operations 
(including deletes) on all replicas. These per-key guarantees follow immediately 
from the properties of the RAFT log and by requiring clients to perform a quorum 
read (logged as an entry in the RAFT log).

In addition, Cyclone also provides clients with a weak read that does not require a 
quorum read thereby requiring only one round trip to a server.  A weak read of a key sees a 
prefix of the linearized sequence of updates to a key that includes all successful update operations made 
by the client before it issues the weak read. To do a weak read, 
clients maintain the last known term of the RAFT leader locally and ensure they always talk
to a RAFT leader that has at least the same term. This ensures that clients see their own writes
avoiding split brain problems where a network partition allows clients to
talk to old RAFT leaders after having committed changes to a new RAFT leader. In addition, a weak 
read waits until all pending RAFT log entries have been committed and applied before executing. 
This ensures that in the event of a failover, any previous committed updates from the same client 
have been applied and are visible to the weak read. A weak read is not linearizable as the RAFT 
leader responding to the weak read might actually no longer be the leader - a new one having been 
elected without it realizing, perhaps due to a network partition. This means that while a weak read 
from a client sees its own writes, in some rare cases it might not see later writes by other clients.

Finally, Cyclone allows operations to multiple keys in an atomic batch, where
either all updates in the batch are applied or no operation in the batch is
applied - a property that holds \emph{at all replicas}, regardless of failures. Batched operations 
are supported in Rocksdb to allow referential integrity between keys - where the value associated
with a  key is a reference to another key. We also use a lightweight batched operation to support 
snapshots - that can be viewed as an atomic no-op to all keys. We discuss how batched operations are 
supported by Cyclone later in the paper.

\section{Storage}
\label{sec:storage}
Storage devices such as flash-based SSDs export a block IO interface. Appending
a small update to the write-ahead log requires a synchronous access to the
SSD. Even with write coalescing using a write cache on the SSD, the round trip
time to a current generation NVMe SSD such as the one we use in this
paper~\cite{ssd_spec}, is of the order of 20 us, limiting the throughput to
under 50K ops/sec and resulting in a high baseline latency at even low load (as
shown in Figure~\ref{fig:problem}). This problem leads many system designers to
under-provision key-value store shards ensuring that it operates at moderate to
high load, using group-commit to batch additions to the log. They therefore pay
a price in latency to ensure the bandwidth to storage is fully utilized.

Cyclone adopts a different strategy by making use of novel memory technologies
to obviate the need to make this tradeoff in the first place. We assume a small
amount of non-volatile memory directly addressable by the CPU, rather than being
placed behind a block oriented IO interface. Log appends are done to the ``NVM
log'' placed in this non-volatile memory. However, directly attached persistent 
memory in the form of NVDIMMs today or new memory technologies in the
future~\cite{pmfs} will be of lesser capacity than traditional (NAND based)
flash available through an
SSD. Providing the same capacity as existing flash based logs for key value
stores therefore requires us to provide a second level of the log
placed on a flash SSD (called flashlog).  Entries are drained from
the NVM log to this flashlog in conveniently large units that are a multiple of
4KB -- the optimum IO unit for flash SSDs. This ensures we get the best possible
throughput from a flash device without paying the price of synchronous IO for
small key-value pairs. At the same time, the amount of non-volatile memory
required is small enough to not add undue cost to the server. This is in contrast to
systems like FARM~\cite{farm} that require both the key value store and write ahead log
to be held entirely in non-volatile memory.

\begin{figure}
  \centering
  \includegraphics[scale=0.6]{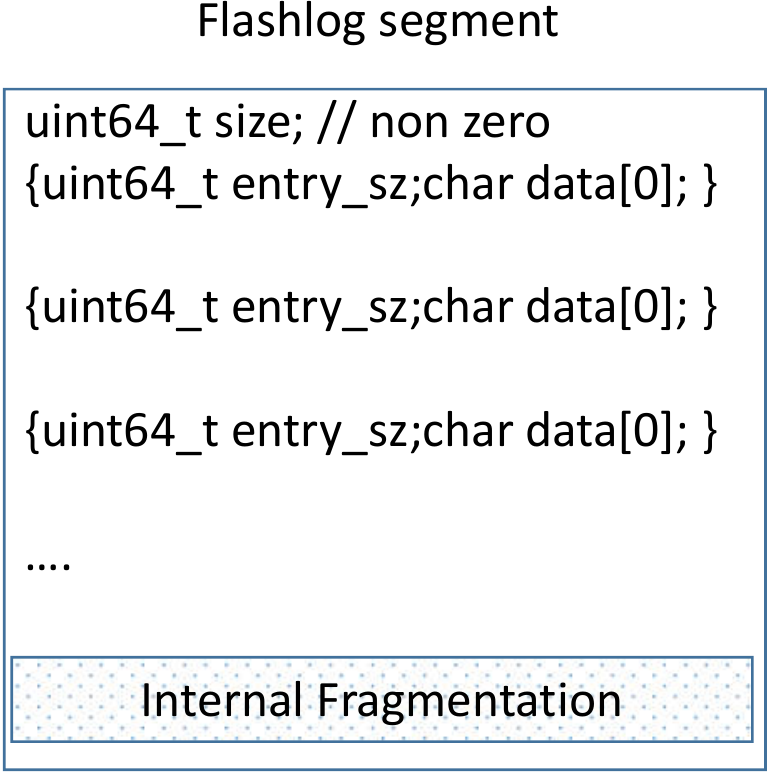}
  \caption{Flashlog segment}
  \label{fig:flashlog_page}
\end{figure}

The flashlog is written out in segments of configurable size (we use 128KB
segments). A segment buffer is prepared in memory (volatile DRAM) using the
layout shown in Figure~\ref{fig:flashlog_page}. We do not allow objects in the
flashlog to cross a 4KB boundary - linking multiple objects together with a
special flag encoded into the size if necessary. We flush log segments out to
the log file using asynchronous direct IO, and therefore we continue to fill log
segment buffers while keeping IO to previously filled buffers outstanding to the
flash drive. We allocate enough buffers to keep a maximum of 32 outstanding
requests to the SSD. To avoid having to do a synchronous metadata flush, we
preallocate (using the posix\_fallocate call) a gigabyte worth of zero filled
disk pages at the end of the log file whenever we hit its end.

In order to recover from a crash, we make two important assumptions about the
underlying SSD. First, we assume that 4KB is the \emph{minimum} atomic unit for
updating pages on the SSD even under power failure i.e. there are no shorn
writes (otherwise known as torn writes) on a 4KB page~\cite{shorn_writes}. We
also assume that the SSD has power loss data protection meaning that writes
cached in the drive's volatile cache are written to the SSD using a backup
capacitor in the event of power failure -- a property of many data center class
SSDs today, including the Intel DC P3600 SSD~\cite{ssd_spec} we use in our
evaluation. Together, these two assumptions mean that we can recover a
consistent prefix of the log on a power failure.

We move log entries from the head of the NVM log to the flashlog buffers in FIFO
order. The NVM log entry is only actually removed when the IO for the
corresponding flashlog page is complete. This means that during recovery we can
have the same log entry both in the NVM log and the flashlog, a condition that
can be detected by examining the log sequence number that we embed in each log
entry.

In this section we demonstrated how Cyclone uses a small amount of non-volatile
memory to improve storage bottlenecks when maintaining a durable log on a single
machine. In the next section, we deal with replicating this log efficiently
across multiple machines.

\section{Network}
\label{sec:network}
Cyclone uses the RAFT~\cite{raft} consensus protocol to replicate the log of
operations across machines. Cyclone provides guarantee to the client that any
operation responded to is durable and its results \emph{available} as long as a
majority of replicas continue to function. Cyclone's failure model is
\emph{fail-recovery}, where nodes that can recover from failure simply continue
to participate in the protocol. This requires RAFT to persist every log entry
\emph{before} it sends it out for replication and before a follower replica
responds to the leader. Storage overheads are therefore not independent of
network overheads, and we must deal with both storage and networking overheads
simultaneously to improve the performance of replication.

We leverage the two level log from the previous section to remove storage
bottlenecks from the critical path of replication in much the same way we did
for logging on a single machine. RAFT is only aware of and replicates the top
level NVM log. Each replica \emph{independently} drains committed (in terms of
RAFT consensus) entries from the NVM log to the flashlog. Block IO therefore is
no longer a consideration when optimizing data movement over the network, unlike
systems that need to deal with such problems~\cite{reflex}. The task of
replication reduces to that of efficiently sending a block of data already
present in directly attached memory over the network. 

The core network operation in replication is to receive a request from the
client at the leader replica and send that exact same request (as a log entry)
to all follower replicas. The leader therefore \emph{multicasts} a received
request packet to follower replicas, a problem well studied in the networking
community when building software packet switches. Software packet switches reach
impressive speeds of millions of packets forwarded per second~\cite{dpdk_perf}, 
a number far in excess of the few thousands of packets we manage in
Figure~\ref{fig:problem}. The key insight in removing network overheads in
Cyclone is therefore to approximate a software packet switch for the networking
component of Cyclone. We do this by implementing Cyclone's network stack on top
of the Data Plane Development Kit (DPDK)~\cite{dpdk}, that provides low latency
userspace access to an ethernet NIC. We discuss Cyclone's network protocol in
Section~\ref{sec:netprot}.

However, the flow of packets through a software packet switch is usually very
simple: the packet enters through an ethernet port and after a simple (usually
stateless) decision is sent out through a set of chosen ports.  In contrast, log
replication requires a complex protocol state machine to decide what to do with
the packet. In addition, unlike software switches we cannot simply forget a
packet after transmitting it. We must append it to the log and act on it after
consensus is reached by a majority quorum. In Section~\ref{sec:dm} we cover the
techniques we use in Cyclone to ameliorate these overheads.

\subsection{DPDK}
\label{sec:netprot}
The Data Plane Development Kit~\cite{dpdk} provides low latency userspace access
to an ethernet NIC, permitting the application to directly send and receive raw
ethernet frames via the transmit and receive queues on the NIC. DPDK is often
used by developers of software packet switches and therefore we leverage
DPDK as a library for building Cyclone.

DPDK does not by itself provide a TCP stack. But this is not a problem, since
RAFT (and indeed most consensus protocols) tolerate network losses and
reordering by design due to the need to support asynchronous communication. In
addition, most datacenter networks rarely drop or reorder packets and provide
full bisection bandwidth between servers that might serve as Cyclone replicas. We
therefore jettisoned TCP and chose to send raw IP packets encapsulated in
ethernet frames. Cyclone takes complete control of an ethernet interface,
receiving all packets directed to it, while the IP and ethernet addresses
provide sufficient information to route the packet if necessary through multiple
switches. We currently follow this communication model both for server to server
communication between replicas as well as client to server communication for
requests and responses. Although a detailed evaluation is made later, switching
from the kernel TCP/IP stack to DPDK reduces the latency between machines in our
testbed from {\tt 18 us} to as low as {\tt 5 us}, providing a significant boost
to performance.

\subsection{Addressing Overheads}
\label{sec:dm}
Software packet switches built on DPDK try to touch as little of the packet data
as possible, minimizing movement of data up and down the cache and memory
hierarchy. Log replication looks like multicast that does not require deep
packet inspection. Software packet switches implementing multicast therefore
simply manipulate packet headers to produce new packets to send on the output
ports. We designed our implementation of the RAFT protocol based on the same
principle.

\begin{figure}
\small
\begin{verbatim}
    event_receive_client_req()
    {
      if(!check_is_leader()) {
        drop_msg
        return
      }
      Prepend raft log term and index
      Persist to local log
      Transmit to follower replicas
    }
\end{verbatim}
\caption{Event handling}
\label{fig:control_plane}
\end{figure}

\begin{figure*}
  \centering
  \includegraphics[scale=0.6]{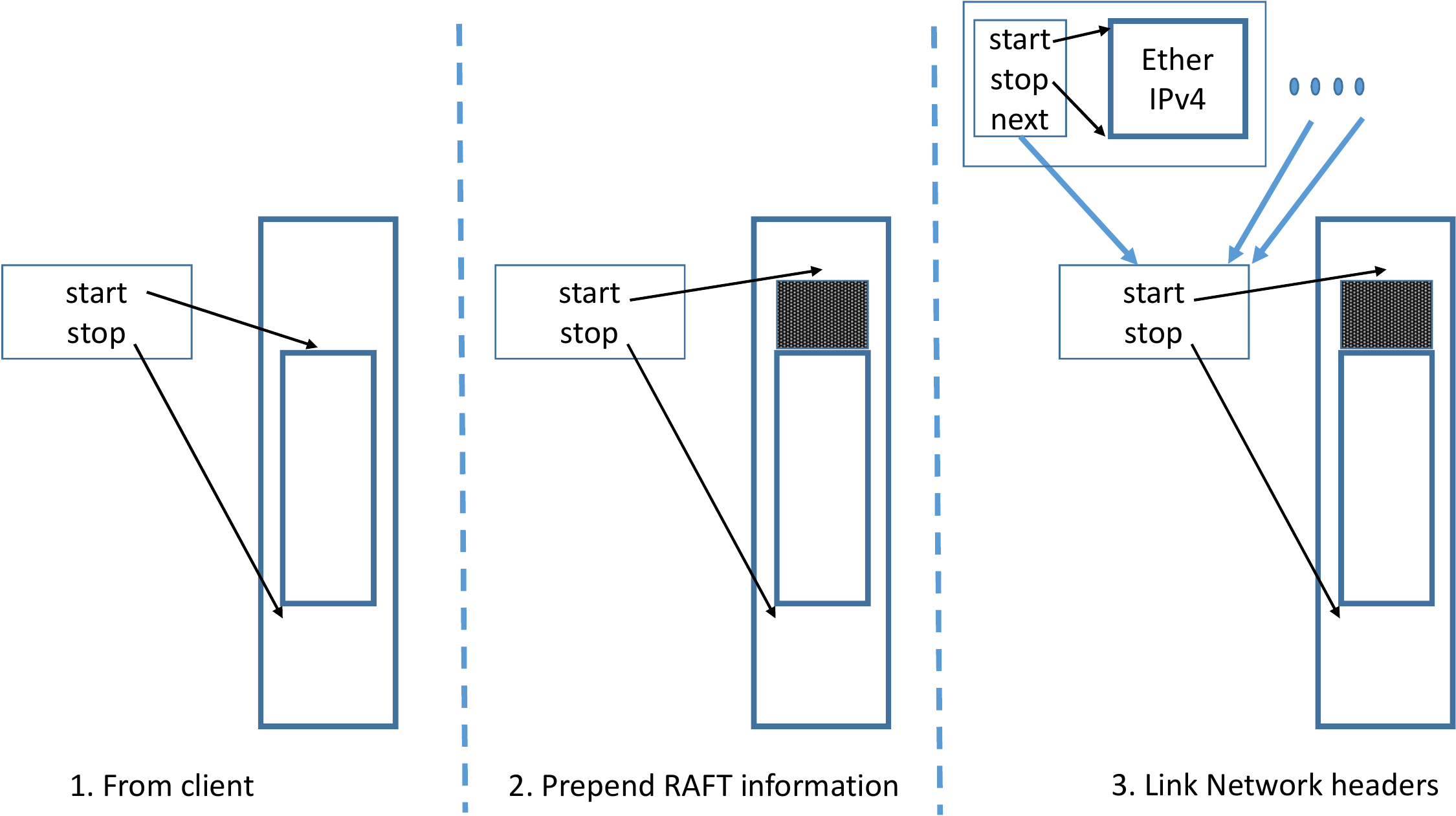}
  \caption{Cyclone network packet layout}
  \label{fig:packet_layout}
\end{figure*}

\begin{figure}
\centering
\includegraphics[scale=0.5]{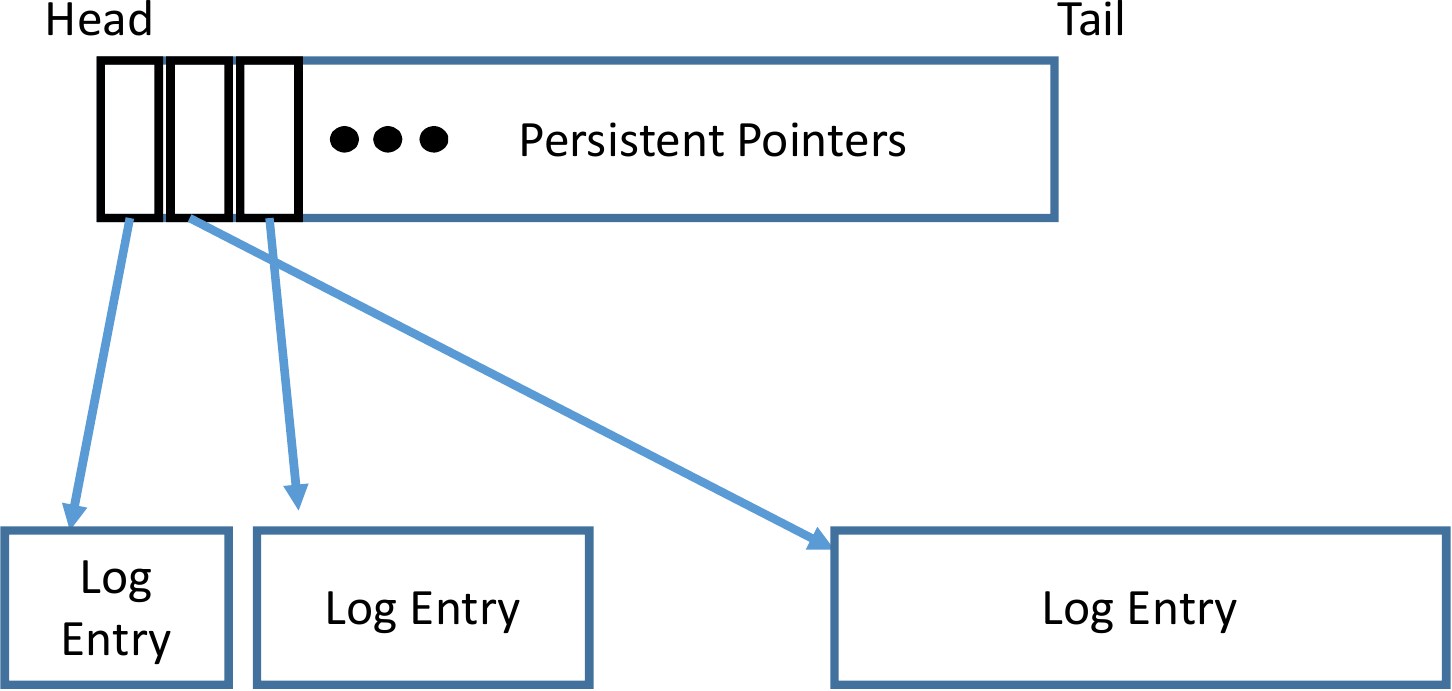}
\caption{NVM log structure}
\label{fig:nvm_log}
\end{figure}

The pseudocode in Figure~\ref{fig:control_plane} describes part of the packet
handling code in Cyclone organized as event handlers triggered on receiving a
packet at the leader. We focus on only one key event for brevity: the event
where a request is received from a client. The first step is to check that this
replica is indeed the RAFT leader in the current view (term). If not, the
message is simply dropped (a timeout causes the client to try a different
server). It then prepends RAFT related information to the packet - this includes
the current term and log index for this entry. Next, it appends a pointer to the
log in Figure~\ref{fig:nvm_log} effectively appending the packet to the
persistent log. Finally, it transmits the packet to follower replicas.

This entire process is done without making any copies of the received packet.
Figure~\ref{fig:packet_layout} illustrates how
Cyclone manipulates packet layouts across the two steps of prepending a RAFT
header and transmitting to follower replicas. DPDK describes packets using an
``mbuf'' data structure. Roughly speaking, an ``mbuf'' consists of a flat array
of bytes actually containing the packet and a fixed size piece of external
metadata that describes various aspects of the packet, most crucially a pointer
to the start and end of the packet in the byte array. DPDK's userspace drivers
receive packets from the NIC such that they are offset in the byte array by a
configurable amount referred to as ``headroom''. We strip off the existing
network headers in the packet and prepend RAFT related information specific to
each log entry in the headroom by shifting the start pointer appropriately.
These operations are standard enough for software packet switches that DPDK
provides convenient library calls for it. For the final step, we need to prepare
the packet for transmission to the various follower replicas. To do this we
prepare a different packet containing a network header for \emph{each}
targeted replica and ``chain'' the data packet to each of these headers. Each
header is then separately handed off to the driver for transmission via the NIC,
carrying the data packet with it by association.

We direct DPDK to use pages backed by NVM for packet buffers. DPDK uses a
concurrent memory allocator based on reference counting - used by both the NIC
driver and CPU cores. This means that packet allocation and deallocation does
not happen in the same order as their corresponding position in the replicated
RAFT log. To deal with this, we use a level of indirection as shown in
Figure~\ref{fig:nvm_log}. The NVM log is maintained as a circular log of fixed
sized pointers to the actual packets. Adding a level of indirection in the NVM
log allows us to separate the FIFO ordered circular log being manipulated by
RAFT from packet data being managed by the memory allocator of DPDK. Both the
circular log and packet data are in NVM. An advantage of this scheme is that it
makes recovery from NVM easy -- appends to the circular pointer log are atomic
and we can use the pointer log to recover allocator state i.e. what pieces of
NVM are currently in use by the log.

RAFT requires that the log entry be persisted \emph{before} it is multicast out
to follower replicas. DPDK userspace NIC drivers operate in DDIO
mode~\cite{ddio} where the packet is directly written into the CPU cache rather
than first being DMAed into DRAM and then fetched by the CPU on demand. We need
to persist the packet to directly attached NVM by executing a cacheline flush
({\tt clflush}) instruction for every cacheline in the packet and the pointer in
the pointer buffer to persist these via the memory bus. This is not too onerous
a burden because we can use the newly introduced {\tt
  clflush-opt}~\cite{clflush_opt} instruction specifically intended to
efficiently flush to persistent memory without the overhead of the serialization
normally introduced by {\tt clflush}. This allows us to hit full memory
bandwidth on current generation platforms, a quantity in excess of 200 Gb/s per
core which is well above the near term speeds of commodity NICs. We execute a
single serializing {\tt sfence} before the sequence of {\tt clflushes} to make
sure any dirty cachelines due to header manipulation related writes from the CPU
are sent to cache.

\begin{figure}
  \centering
  \includegraphics[scale=0.5]{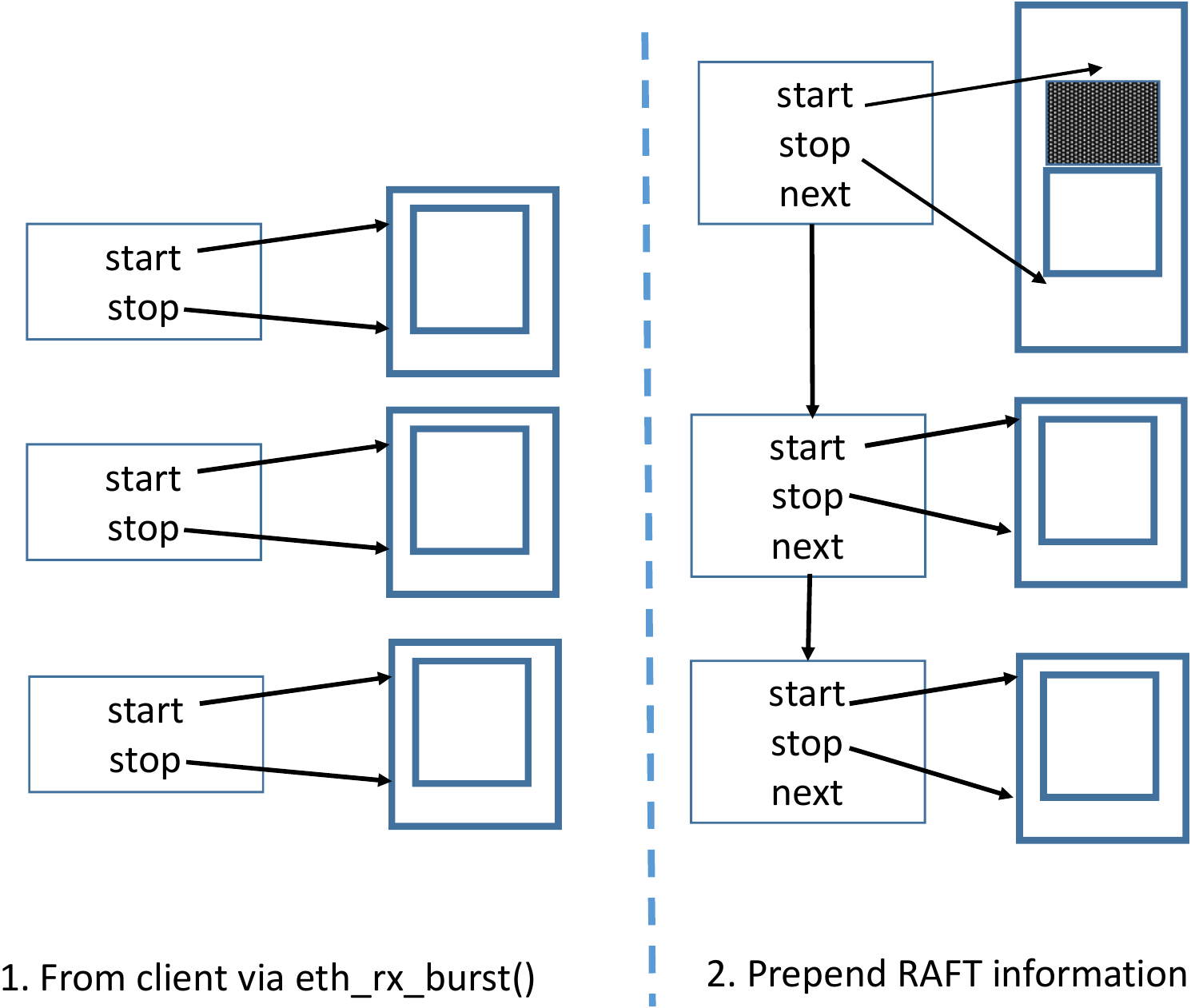}
  \caption{Batching}
  \label{fig:batching}
\end{figure}

Although RAFT is an efficient consensus protocol in the common case, the
protocol state machine still adds significant overhead to each packet, relative
to the time for the packet to flow in from the NIC and back out to the replicas.
We address this problem for the \emph{loaded} case using batching - treating a
whole sequence of client commands as a single RAFT log entry, while avoiding any
copies to group these packets together. Figure~\ref{fig:batching} illustrates
how this is done. We use a burst receive call available in the DPDK userspace
driver to receive a burst of client packets at a time. We then chain these
packets together and treat them as a single log entry from the perspective of
RAFT, amortizing the control plane overheads over the packets (at most 32 at a
time due to current driver limitations). Crucially, batching in Cyclone does not
involve a latency-throughput tradeoff like in many other
systems~\cite{ix-dataplane}. The batch receive call we use in DPDK returns
immediately with whatever number of packets is available, including zero. We
always flush the transmit buffer after every call to DPDK to transmit packets to
replicas. Therefore, we never tradeoff latency for throughput when batching. We
also receive log entries in batches at follower replicas and return a single
acknowledgment for the entire batch, speeding up the progress of the protocol.

Our choice of RAFT as a consensus protocol is driven by the need to efficiently
maintain the log datastructure. Unlike alternatives such as
MultiPaxos~\cite{multipaxos} or Quorum based replication~\cite{quorum}, the
leader is guaranteed to have the most up to date logs. There is therefore no
case where a leader needs to receive log entries to ``fill'' holes in its logs
from follower replicas, simplifying the protocol state machine we need to
implement. A leader can immediately reject any responses that are not at least
as current as its view (term). In turn, the persistent log operates as a double
ended queue, with entries either being appended (or possibly deleted at
followers) at the end and being deleted from the front (after commit). This is
critical to efficient operation of Cyclone, as it allows a simple top level
structure (a circular buffer of pointers) to represent the log.

\section{Parallelism}
\label{sec:parallelism}

%\begin{figure}
%  \centering
%  \includegraphics[scale=0.35]{race.pdf}
%  \caption{Race with multiple physical logs}
%  \label{fig:race}
%\end{figure}

Although the work described in the previous section significantly boosts the
network performance of replication in Cyclone, it still does not come close to
saturating the capabilities of even the commodity NICs that we use. This is
because NICs today encapsulate significant parallelism in terms of
multiple send and receive queue pairs. Exploiting this capability to improve
throughput requires us to remove the bottleneck of a single sequential log in
RAFT. To this end, we extend the implementation described thus far to run
multiple copies of the RAFT consensus protocol each maintaining and replicating
its own two level log. We refer to these logs as physical logs. All these
instances however exist in the same process address space as the key value store
application itself and therefore manipulate the same application. The number of 
instances is a fixed property of the Cyclone service and cannot be changed after 
startup.

The first question we deal with is -- how do Cyclone clients decide which
physical log to send a request to? The guarantees from Cyclone (except batched
writes)  cover ordering of updates to a single key (Section~\ref{sec:sysarch}). They can 
be satisfied by ensuring that all reads and writes to a key go to the same physical log
(i.e. RAFT instance). We achieve this by hashing the key to select the physical
log.

Multiple physical logs in Cyclone operate in a shared nothing manner by
partitioning the NVM and SSD space evenly between them and by allocating
dedicated NIC queue pairs to each instance of RAFT. This works well because the
memory hierarchy including the NVM and the SSD efficiently support concurrent
operations. The only synchronization necessary is when doing reads or writes to
the single shared key value store.  The level of concurrency therefore is
constrained only by the concurrency available in the software architecture of
the persistent key value store itself, for which good designs
exist~\cite{flodb}.

\subsection{Ganged Operations}
Using multiple instances of RAFT leads to a serious problem with batched operations, 
since we need to split up the batch into a mini-batch for each of the physical logs. There
is however no guarantee that the request for a mini-batch will succeed breaking the atomicity
requirement for batched updates. The obvious and simple solution is to do two phase commit across
the logs, ensuring all or nothing semantics. However this was unacceptable to some customers who
pointed out that ensuring that an external co-ordinator implementing two phase commit does not itself
fail requires running a separate replicated service - with its attendant resource and reliability headaches.
We were therefore challenged to come up with a solution that did not require two phase commit.

Our final solution is based on two observations. First, we \emph {do not} require that either all mini-batches
icommit to their raft logs or none do, What we actually requires is that either all are \emph{applied} to Rocksdb or 
none of them are. Put another way we should not apply a mini-batch from a ganged operation if any of the other mini-batches
have failed to commit to their RAFT log. This means that the client can be stateless. The second observation is that
checking whether the other minibatches have committed is easy because the different RAFT instances can communicate 
through shared memory on the same machines unlike the more general case of distributed participants in two phase 
commit.

We now describe our solution called ``ganged operations" in Cyclone. 
The first key problem we need to deal with is that the leaders for the different
RAFT instances to which the keys map can be located on different machines. To
avoid having a client co-ordinate different machines for a single operation, we
constrain the RAFT leader election algorithm for the different RAFT instances to
converge to the same machine. We do this by triggering re-elections for a RAFT
instance other than that of the first physical log until it is on the same
machine as the \emph{first} physical log. This process is therefore resilient to
failed machines. RAFT requires that the leader have the most up to date log. To
ensure that the process converges we therefore ensure that the currently elected
leader brings a majority quorum up to date before triggering a reelection. We
also do not accept any requests from clients until all leaders are
co-located. We assume that no network partitions occur \emph{within a machine}
i.e. some RAFT instances are able to send and receive packets, while others are
not, a possibility we discount due to our single process design.

Next, we need to simultaneously inject the ganged operation into all
participating physical logs on the machine hosting the leaders. Clients always
dispatch batched writes to a fixed co-ordinating physical log/RAFT instance
(henceforth called the co-ordinator), which is then responsible for forwarding
the request to the participating logs. We make use of packet cloning primitives
to avoid making physical copies of the packet, generating indirect references
instead. The co-ordinator also adds a ``nonce'' - a unique timestamp to the
packet. In addition, the co-ordinator adds a unique view number to the packet
containing the term numbers of all the participating RAFT instances (read from
shared memory of the co-located leaders).  The event that applies a ganged
operation is described in the pseudocode of Figure~\ref{fig:ganged_ops}. We
defer the discussion of how we generate the nonce to later in this section. We
also assume that a unique barrier is allocated in shared memory for each ganged
operation. We discuss later how this is done without using dynamic allocation.

The complexity in Figure~\ref{fig:ganged_ops} arises from the need to handle
both shared memory concurrency and failure during replication. Without failure,
Figure~\ref{fig:ganged_ops} is straightforward. A barrier is executed on all
RAFT instances. Once replication is complete on all participating physical logs,
a state indicated by all necessary bits being set in the barrier mask, the RAFT
instances simultaneously execute the ganged operation.

A failure causes a new leader to be elected for the affected physical log,
moving forward the term. If the participants of a ganged operation do not
detect this case, they could be left waiting forever. We detect failure on any
RAFT instance by having each instance publish its current term and continuously
comparing the view in the ganged operation to it. If any RAFT instance has moved
past that term, the ganged operation is then terminated.

\begin{figure*}
  \centering
  \small
  \bf
\begin{verbatim}
    // Apply ganged operation
    event_apply_ganged_op (packet, barrier)
    {
      if(co-ordinator)
        atomic set bit me in barrier.mask         
        do
         for each participant p in operation
          if public_data[p].view > packet.view
           barrier.failed = true 
           atomic set bit for p in barrier.mask
        while barrier.mask != mask of all participants
        if barrier.failed
         send retry to client
        else       
         execute operation
         send response
      else
        wait until 
         public_data[co-ordinator].view > packet.view OR
         barrier.mask == mask of participating cores
        if public_data[co-ordinator].view > packet.view OR
           barrier.failed
         send retry to client
        else
         execute operation
    }
\end{verbatim}
\caption{Ganged Operation}
\label{fig:ganged_ops}
\end{figure*}

The assumption in Figure~\ref{fig:ganged_ops} is that each ganged operation is
mapped to a unique barrier. We achieve this by using a fixed piece of memory
owned by the co-ordinator to hold the barrier and write the nonce to it in order
to indicate that the barrier is active for the corresponding ganged
operation. Participants watch for the nonce to know when to execute the
ganged operation barrier in Figure~\ref{fig:ganged_ops}, while also monitoring
the leader's published view to detect the case where the ganged operation fails
to replicate on the co-ordinator's physical log.

Finally, we describe how we generate the nonce. The nonce is generated on the
co-ordinator RAFT instance by concatenating the ethernet MAC ID of the first NIC
on the system with a 64 bit value that is the number of CPU timestamp counter
cycles since epoch time (read from the real time clock at startup plus the
number of cycles from the CPU {\tt rdtsc} instruction). The nonce can only be
repeated if the same machine manages to fail and come back up in less time than
the real time clock drift (controlled with NTP), a possibility that we discount.

Ganged operations possibly constitute the most complex part of Cyclone but the
code weighs in at well under a couple of hundred lines. We believe that
this additional complexity is still small compared to distributed transactions
using two-phase commit - especially when taking into account failure recovery -
as used in systems such as FARM~\cite{farm}.

\section{Evaluation}
\label{sec:evaluation}
We evaluate Cyclone on a 12 node x86 Xeon cluster connected via a 10 GigE
switch. Three of the machines are equipped with 1.6TB Intel DC P3600 SSDs and
4*10 GigE ports. The remaining nine machines do not have SSDs and have only one
10 GigE port, serving as clients for most of the experiments. We turn on jumbo
frame support in the 10 GigE switch to enable maximum use of batching in
Cyclone. As with other work~\cite{faast}, we use DRAM on the machines to proxy
for NVDIMMs where necessary - the NVM needed never exceeds 64 MB
regardless of the size of the key value store or second level log on flash. We
divide the evaluation into three parts.

First, we evaluate Cyclone's performance with a single level log as a pure
software packet switch. For this purpose, Cyclone uses a dummy server stub that
simply echoes the client request back to it. Next, we evaluate
performance with the dummy stub, but with the addition of the second level of log
on flash. Finally, we evaluate performance when integrated with
Rocksdb~\cite{rocksdb} as an alternative to Rocksdb's write ahead log.

Unless otherwise mentioned, we use a 60 byte header followed by an optional
payload for experiments. We log both the header and payload. Also, unless
otherwise mentioned we use 8 physical logs (and associated RAFT instances) each
mapped to a dedicated core. The remaining cores are dedicated to the stub
server or Rocksdb, as the case may be.

\begin{figure}
\includegraphics[scale=0.6]{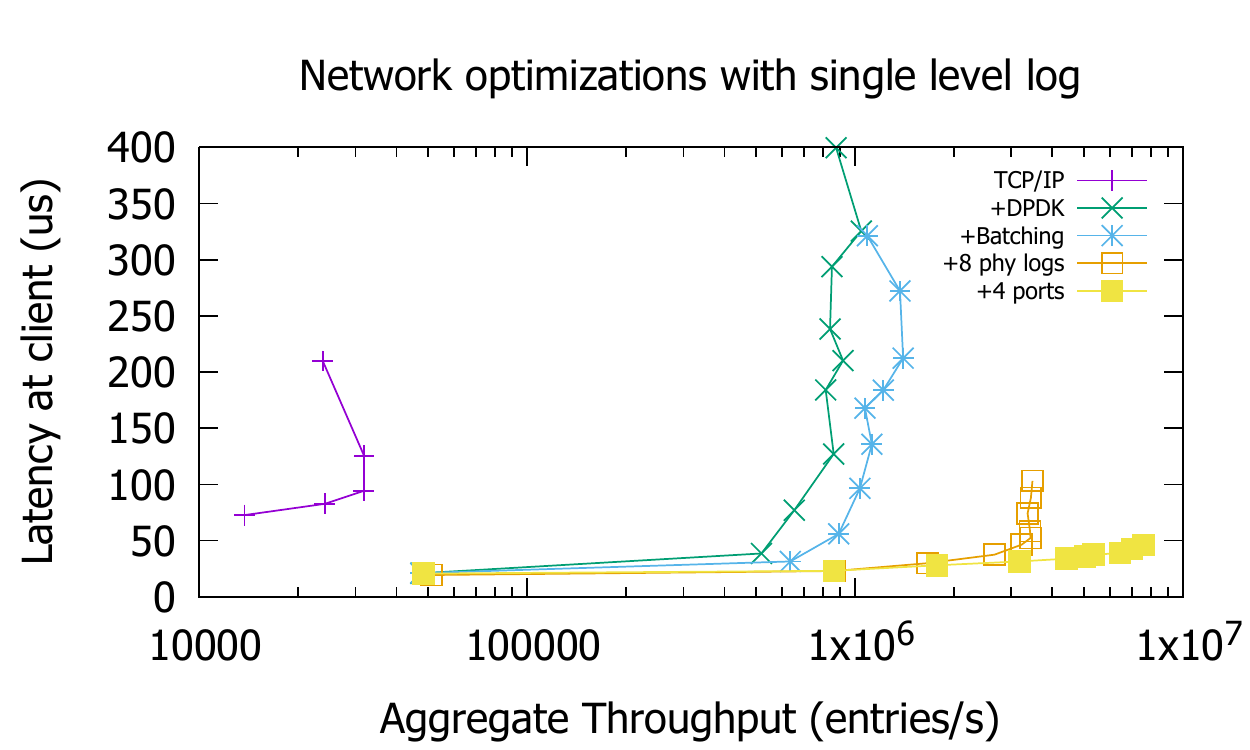}
\caption{Network optimizations for top level log}
\label{fig:network_opts}
\end{figure}

We begin by systematically evaluating network stack related optimizations
applied in Cyclone to replicate the NVM log in Figure~\ref{fig:network_opts} -
with no payload. The y-axis reports latency seen at the client (which means two
network round trips with replication). Using TCP/IP to replicate a RAFT log tops
out at around 30K entries/s.  Switching to DPDK (the line marked +DPDK) improves
the throughput by an order of magnitude to around 500K entries/s. Using batching
(the line marked +batching) improves the performance further bringing us close
to a million entries/s. Scaling to 8 physical logs (+8 phy logs) improves
performance to close to 2M entries/s. Finally using all 4 ethernet ports on the
machine to replicate entries improves performance considerably to 6M
entries/s. In all, performance improves by 200X over the TCP/IP single log
baseline. Cyclone also considerably improves the latency for replication, from
close to 100us with TCP/IP to around 30us at peak throughput. 

One can draw three important conclusions from
Figure~\ref{fig:network_opts}. First, one can indeed treat log replication as a
software packet switching problem \emph{provided} the persistent log can be held
in directly attached memory eliminating the block interface from the critical
path. Second, relieving the CPU from the overhead of data copies has a
significant positive effect on performance. Second, using multiple physical
logs is essential to exploiting the concurrency available at the level of the
NIC (even a single one!) that would otherwise go wasted due
to the serializing abstraction of a single log. Finally, we note that Cyclone
achieves about 50\% of the line rate across the four ethernet
ports. This is essentially the cost of running a consensus protocol in
software.

\begin{figure}
  \includegraphics[scale=0.6]{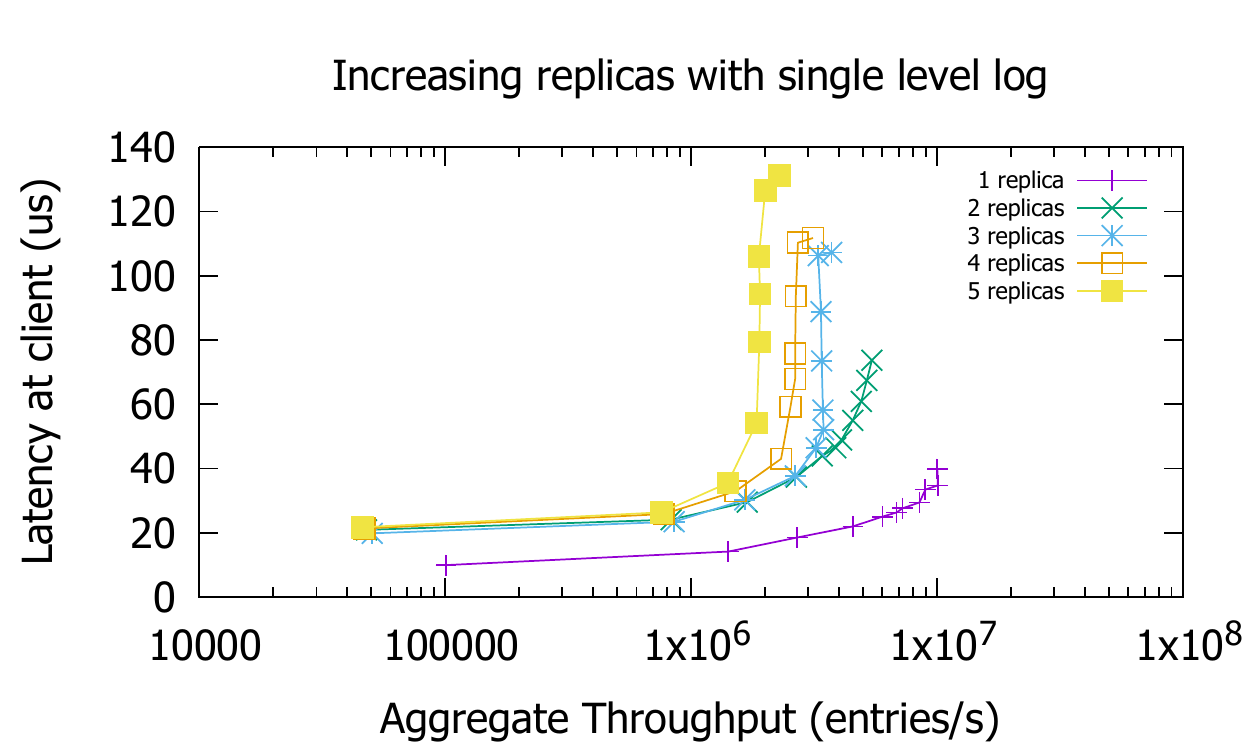}
  \caption{Impact of replica count}
  \label{fig:replicas}
\end{figure}

\begin{figure}
  \includegraphics[scale=0.6]{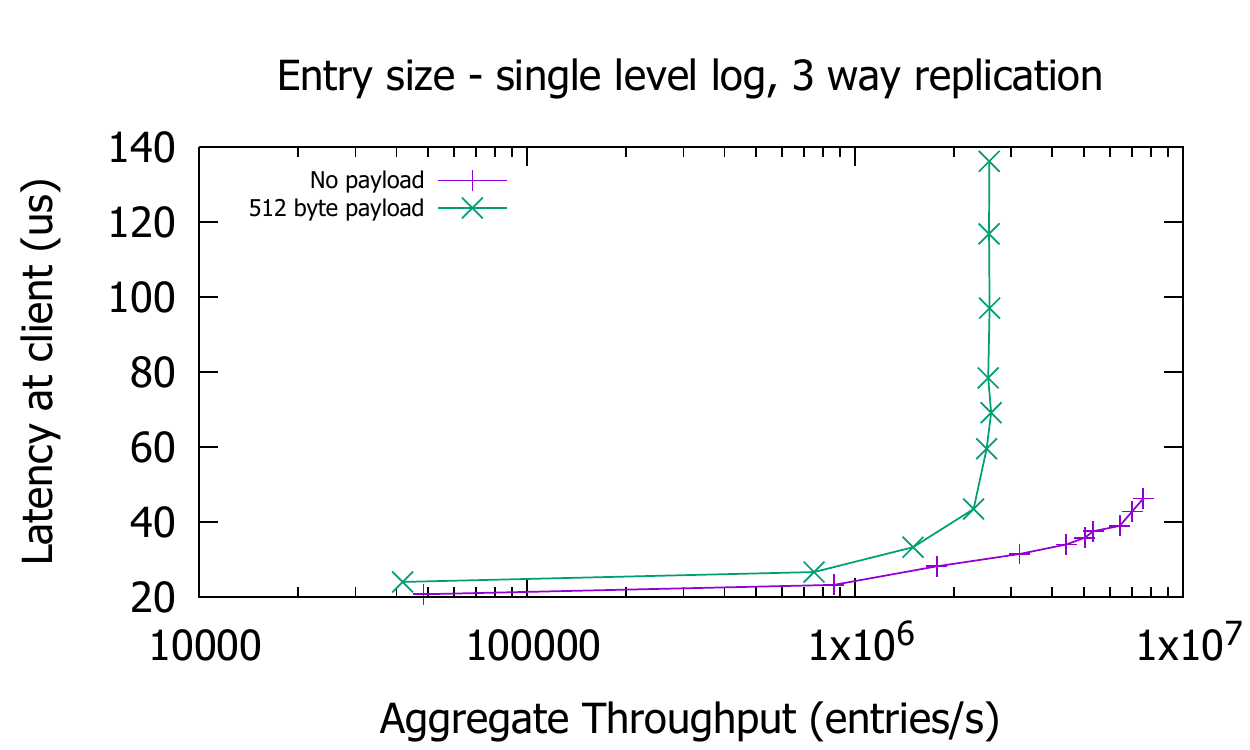}
  \caption{Impact of payload size}
  \label{fig:payload}
\end{figure}

Both the replica count and payload size can have significant impact on Cyclone's
network performance. First, the number of replicas dictates the outgoing message
rate from the leader replica and therefore increasing the replication factor can
decrease Cyclone's performance. Figure~\ref{fig:replicas} shows the impact of
varying replica count. Using only a single replica cuts out a network round trip
and shows the best unloaded latency (10 us) and peak throughput (near 10M
entries/s). Adding replicas decreases the peak throughput down to around 2M
entries/s with 5 replicas. We note that a number of previous pieces of
work~\cite{faast, farm} use three replicas and therefore we focus on three
replicas for the replicated cases we consider below. The second factor that
dictates Cyclone's performance is the size of the log entry being
replicated. Figure~\ref{fig:payload} shows the effect of increasing the payload
size from zero to 512 bytes. Peak throughput drops from 6M entries/s to
approximately 2M entries/s. At this replication rate, the leader replica needs
to transmit data at approximately 30 Gbit/s. Coupled with the cost of network
headers all four 10 GigE ports are now saturated and therefore Cyclone hits the
network line rate bottleneck at this point. It is worthwhile to compare this to
the case with small updates, where running the consensus protocol was a
bottleneck to reaching line rate.

\begin{figure}
  \includegraphics[scale=0.6]{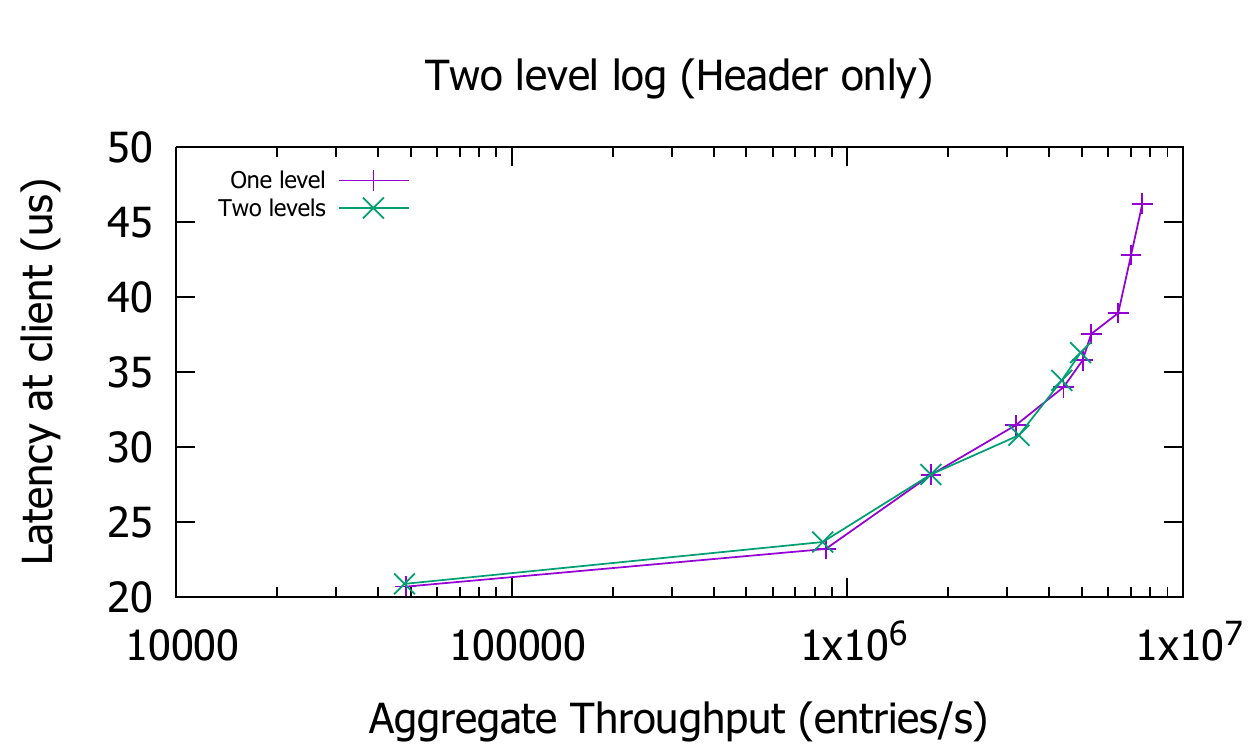}
  \caption{Adding second level log}
  \label{fig:flashlog}
\end{figure}

\begin{figure}
  \includegraphics[scale=0.6]{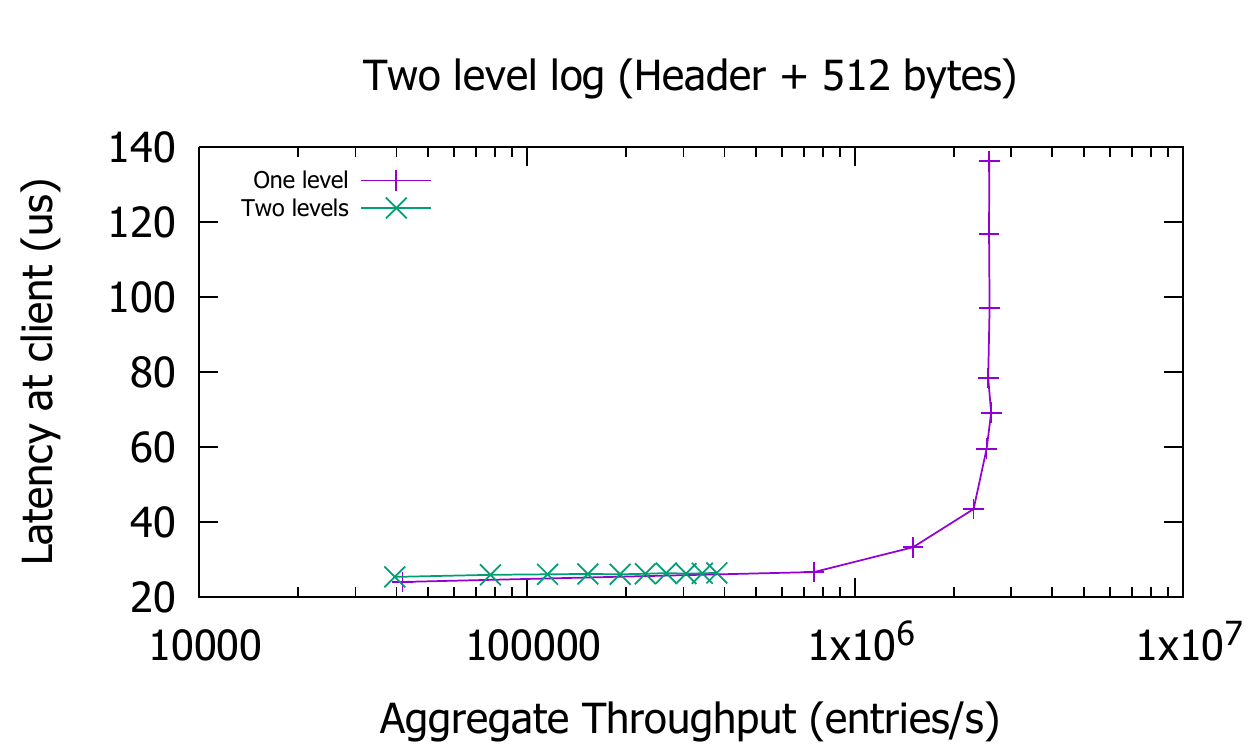}
  \caption{Second level log - 512 bytes payload}
  \label{fig:flashlog_512}
\end{figure}

We now turn our attention from the network component of Cyclone to the storage
one by adding the second level flashlog. We evaluate the impact of adding the
flashlog in Figures~\ref{fig:flashlog} and \ref{fig:flashlog_512}. The benefit
of the second level flashlog is that it lets us keep the same amount of log
space as with the flash based implementation, and not limit it to the amount of
NVM available in the system.
The hypothesis was that batching the movement of data from the
NVM log to the flashlog would be sufficient to hide the latency of the block IO
device from network speed replication. The results confirm that in terms of
latency for the low to moderate load cases, our two level log arrangement is
effective at hiding the latency of the NVMe SSD. The picture however is
different for peak throughput. For small updates there is no impact on
throughput when we add the SSD. On the other hand 
Figure~\ref{fig:flashlog_512} shows that using a 512 byte payload has a
significant impact on peak throughput - it drops to approximately 350K
ops/sec. This corresponds to around 50K 4KB IOPS to the SSD to write out the
flashlog pages. This is in fact the expected IOPS limit with the 32 outstanding
requests to the drive that we maintain (Section~\ref{sec:storage}).
A final observation on Figure~\ref{fig:flashlog_512} is that once we are past
the storage bottleneck the latency spike is dramatic and large enough to trigger
Cyclone's failure detector and repeated retries from the clients. There are
therefore no points on the ``knee'' of the curve as in the pure packet
switched one-level log case.

One can therefore conclude that using a small amount of directly attached NVM is
effective at hiding the latency of a background block storage device from the
critical path of replication. Further, for small updates the
bottleneck is the capability of the software to run the consensus protocol, but
the bottleneck shifts to the secondary storage device for larger request sizes.

\begin{figure}
  \includegraphics[scale=0.6]{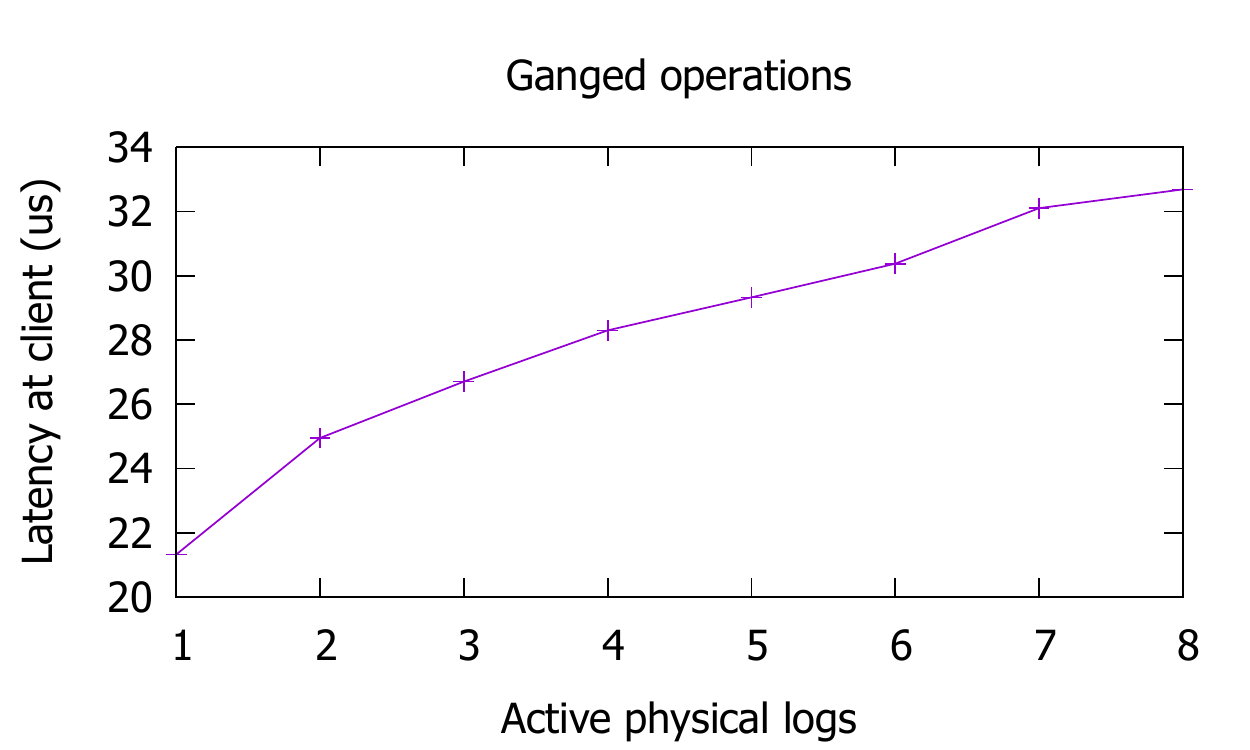}
  \caption{Ganged Operations}
  \label{fig:ganged}
\end{figure}

We now evaluate ganged operations. The primary purpose of
ganged operations is to avoid the need for distributed transactions to
manipulate what is a single shared memory image and therefore we were most
concerned about unloaded latency given the complexity of synchronizing different
replication quorums as well executing our rendezvous protocol on multiple
cores. We therefore setup an experiment where a single client -- reflecting the
unloaded case -- made ganged requests to the replicas. We varied the number of
physical logs participating in the request from one to the full complement of 8.
Figure~\ref{fig:ganged} shows the results. Unloaded
latency increases slowly as we increase the number of active physical logs - to
around 32 us from the baseline of 21 us. The reason for this increase is that
the DPDK userspace NIC driver pipelines request processing for communication
between the CPU core and the NIC. Simultaneous replication on multiple physical
logs represents a worst case for this arrangement as all eight cores try to
send replication messages to their replicas, adding about a microsecond of
serial latency each, to dispatch requests to the NIC. Regardless, the experiment
underlines the value of our ganged operation design over a distributed
transaction that would need 8 network hops for the two phase commit (and
replication) using an external co-ordinator, a minimum of 80us.

\begin{figure}
  \includegraphics[scale=0.6]{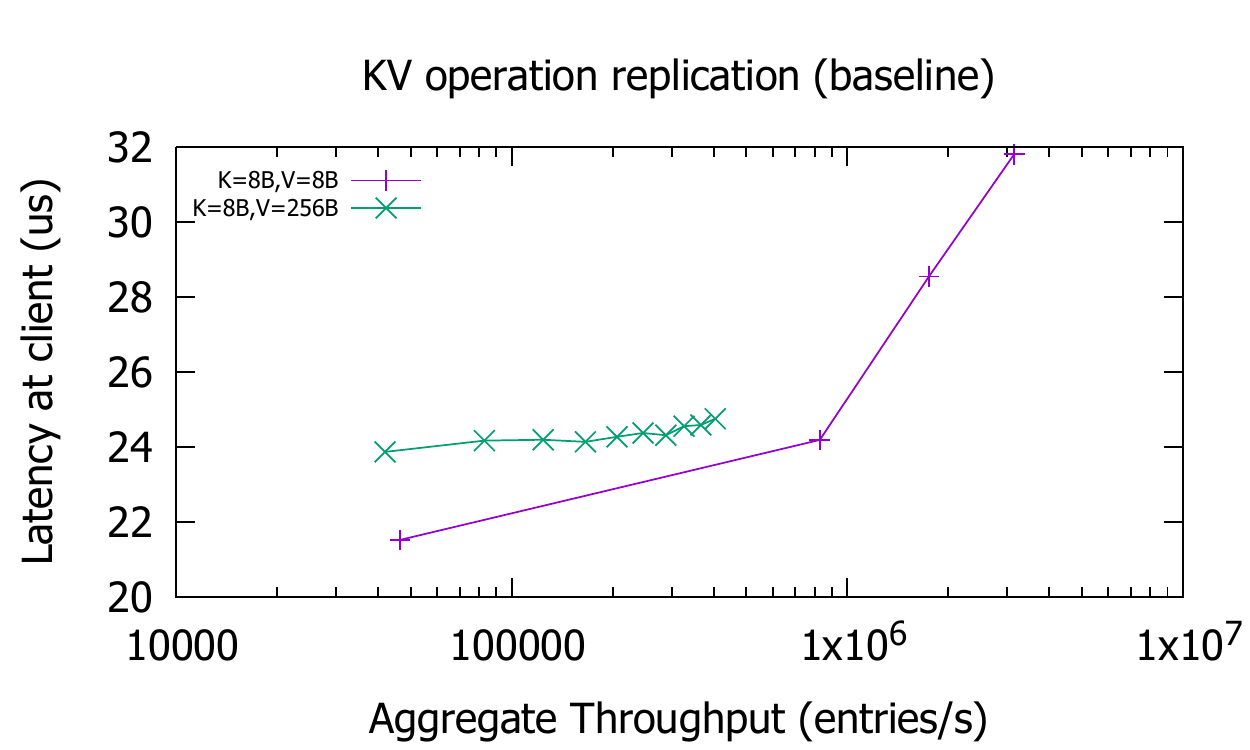}
  \caption{Baseline KV replication}
  \label{fig:kv_baseline}
\end{figure}

Next, we evaluate Cyclone integrated with the Rocksdb persistent key value
store. Rocksdb is a complex commercial grade persistent key value store and
an important target for practical use of Cyclone.  Rocksdb is accompanied by a
complex array of performance tuning knobs to get the best performance from
flash. Designing LSM tree based key value stores to extract good performance
from flash and make best use of DRAM on the machine is an area of active
research~\cite{flodb, triad}. One of our main goals is to demonstrate that
our conclusions will apply even as Rocksdb performance continues to improve
with the integration of new ideas and better SSDs. Keeping this in mind, we
configured RocksDB to place all files for the
key value store (SSTables) on a RAMdisk, which presumably represents the limit
in performance for both software enhancements as well as improvements to
flash-based SSDs. However, both Rocksdb's own write ahead log and the
alternative of Cyclone's second level flashlog are placed on the SSDs. In
effect we make availability a harder problem in this setting to ensure that our
design is future-proof.

For the first experiment with Rocksdb we use 8 byte keys with either
8 byte values or 256 byte values.  Since our
focus is replication and only update requests are replicated, we use a 100\%
update workload to test the capability of the system. This involves loading 100
million key value pairs and during the test updating the value associated with a
key picked at random.

Before evaluating with Rocksdb, we measure the baseline performance of
replicating the log with Cyclone for the given request sizes - Rocksdb performs
a no-op. We note that in addition to the key and value, we are also logging
RocksDB specific request data such as operation type and the request header,
making this different from the previous experiments.
Figure~\ref{fig:kv_baseline} shows the baseline performance for the chosen
request sizes. With the smaller request size, Cyclone can conservatively sustain
close to a million requests a second at a latency of just under 25us. With the
larger request size, Cyclone can sustain around 350K requests a second, again at
a latency of just under 25 us. Armed with these baseline numbers we now examine
how well Cyclone performs with Rocksdb.

\begin{figure}
    \centering
    \includegraphics[scale=0.6]{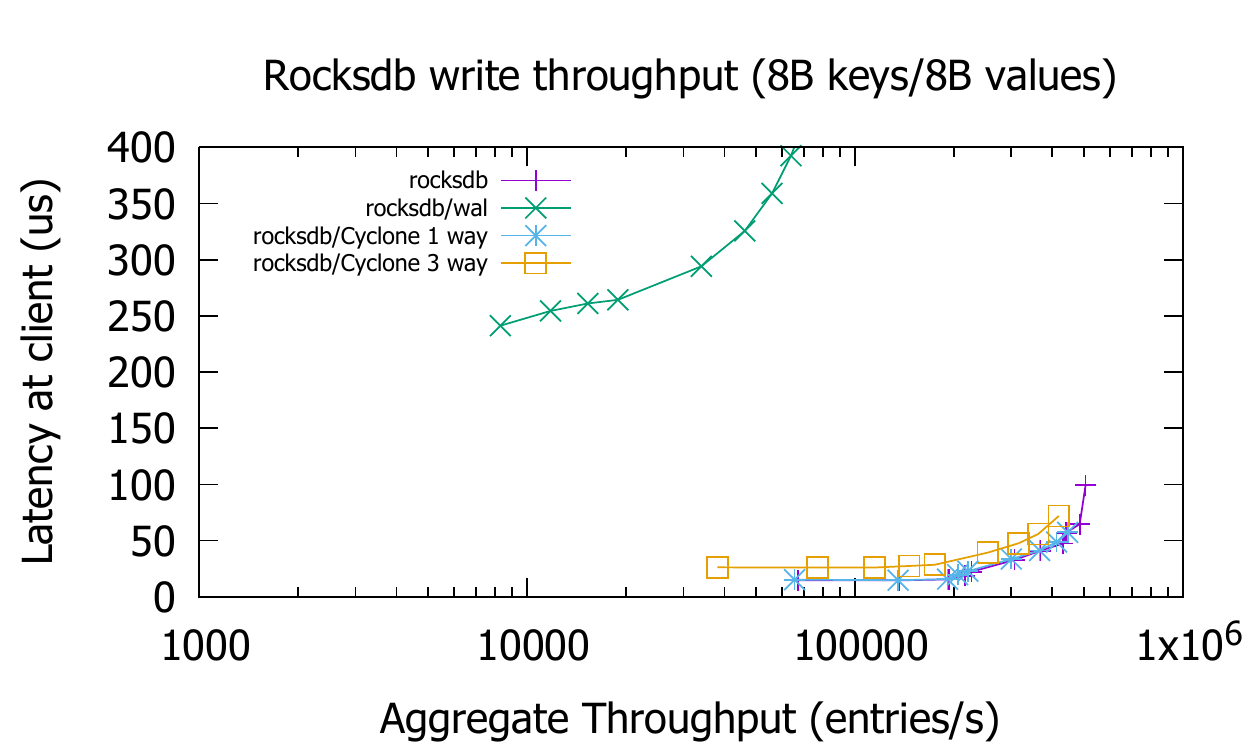}
    \caption{Rocksdb - small updates}
    \label{fig:rocksdb}
  \end{figure}

\begin{figure}
    \includegraphics[scale=0.6]{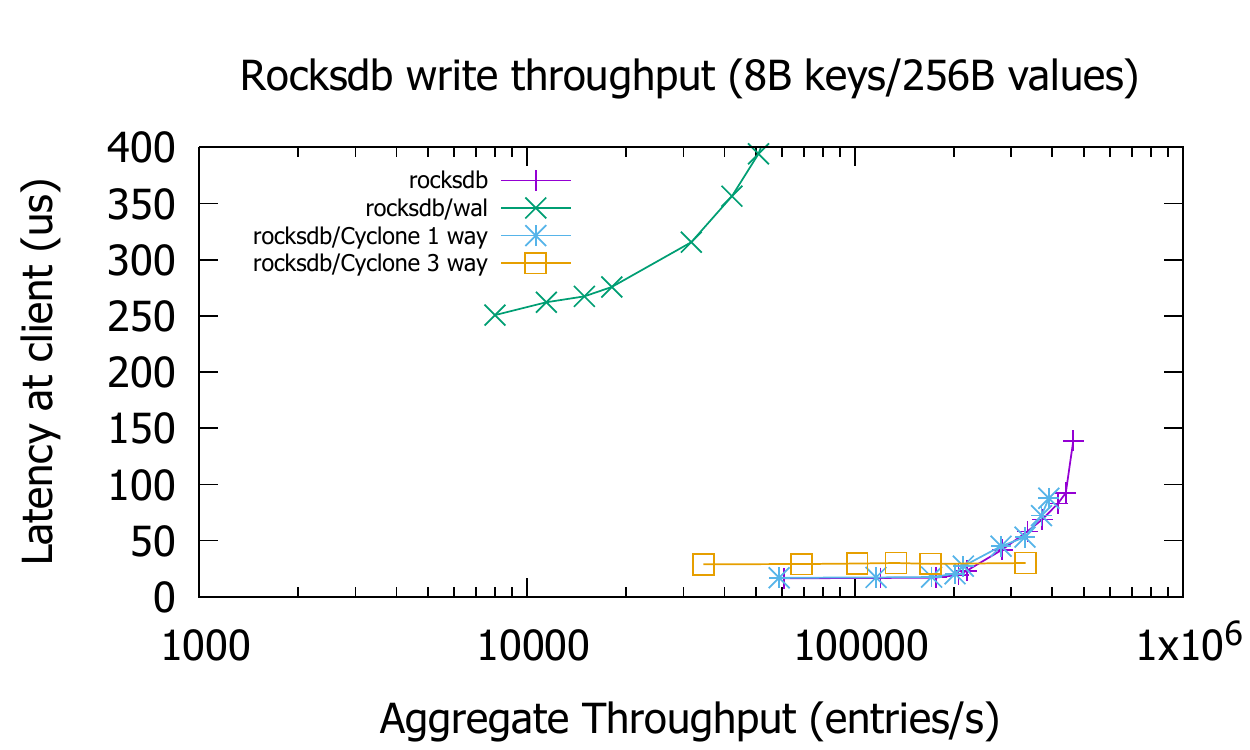}
    \caption{Rocksdb - large updates}
    \label{fig:rocksdb_256}
  \end{figure}

The performance of Rocksdb with Cyclone for the small update workload is shown
in Figure~\ref{fig:rocksdb} - essentially presenting the solution to the problem
demonstrated in Figure~\ref{fig:problem}.  We consider four different
setups. The line labeled 'rocksdb' is the key value store running with no
logging whatsoever - a system crash would lead to data loss. The line labeled
'rocksdb/wal' is for Rocksdb running with its write ahead logging turned on. The
large gap between these two is the overhead of the existing Rocksdb WAL
solution. The line labeled 'rocksdb/Cyclone 1 way' is a two level Cyclone log
but without any replication. The line almost exactly tracks the performance of
Rocksdb. Cyclone is able to provide a write ahead log with no overhead to
Rocksdb. The line labeled 'rocksdb/Cyclone 3 way' is with 3-way replication
turned on. Other than a 20us delta due to the extra network round trip, the line
almost exactly tracks Rocksdb performance with no logging. Cyclone therefore
provides high availability to Rocksdb at a fraction of the cost of its existing
single machine write ahead log. We also repeat the experiment for the larger
update size in Figure~\ref{fig:rocksdb_256}. The conclusions are identical:
Cyclone solves Rocksdb's write ahead logging problem.

  \begin{figure}
    \includegraphics[scale=0.6]{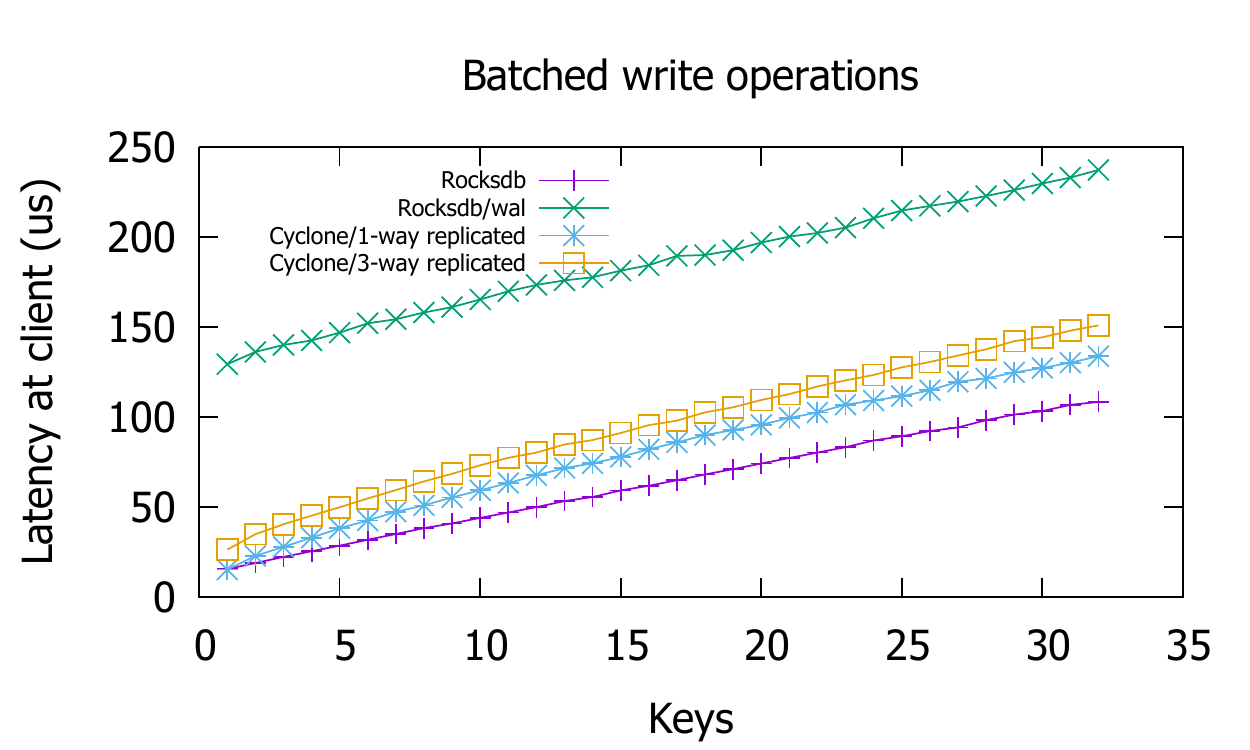}
    \caption{Rocksdb - batched writes}
    \label{fig:rocksdb_multi}
\end{figure}

Next, we consider the problem of supporting Rocksdb's write-batch operation that
atomically writes a set of key-value pairs into the KV store. This is a
practical application of ganged operations in Cyclone. In order to
perform the operation with Cyclone managing the log, the client must issue a
ganged operation across physical logs owning the keys in the write batch.
A key concern here was whether Cyclone would add any latency to the
operation due to the extra synchronization needed across replication quorums (
shown in a previous experiment).
We examine the problem for the unloaded case
and small updates in Figure~\ref{fig:rocksdb_multi}, for a single client with
increasing number of keys in the batch (up to 32 keys).
The line labeled Rocksdb is with no write ahead logging. We note an increasing
latency for this baseline indicating Rocksdb itself takes longer
with larger key batches. The existing option of Rocksdb/wal has considerably
larger latency. Cyclone does an effective job of cutting down on this latency
even as it needs to pay a price for synchronizing multiple physical
logs. Cyclone therefore provides effective replicated logging for batched
operations using the idea of ganged operations, while scaling the performance of
single operations using its multiple physical logs.

\begin{figure}
\includegraphics[scale=0.6]{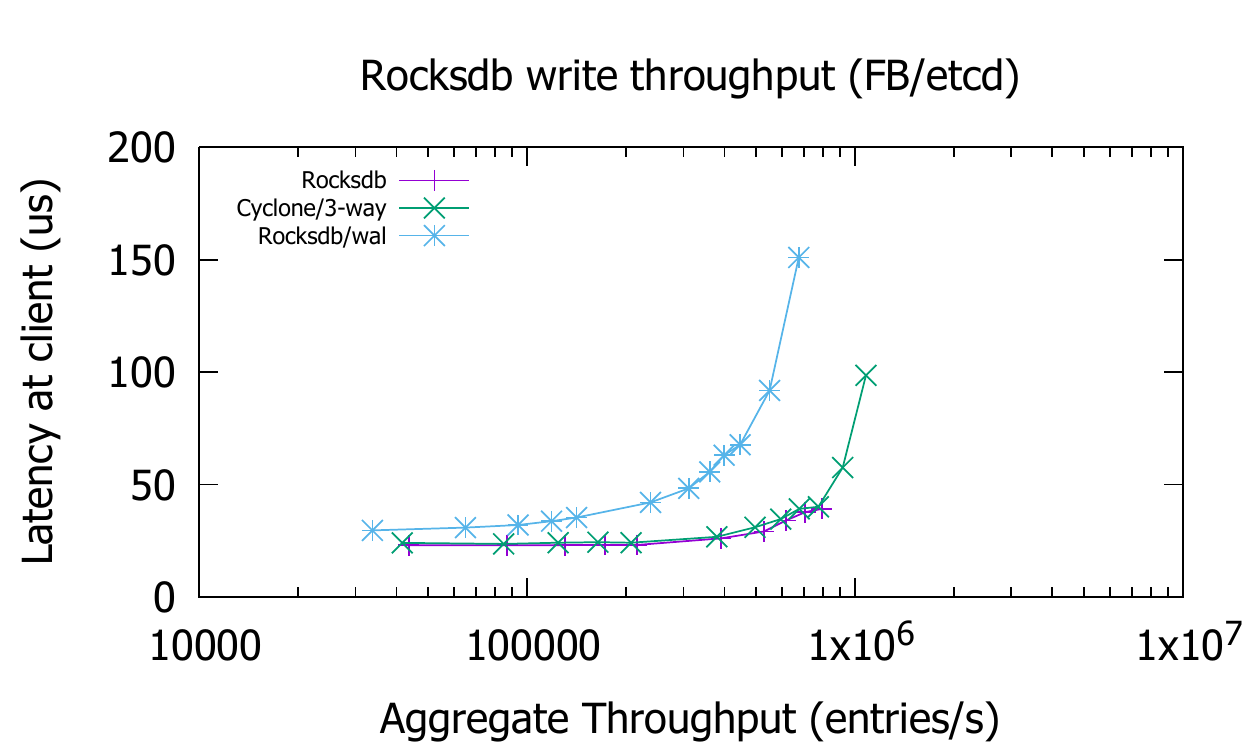}
\caption{Rocksdb - read heavy}
\label{fig:fb}
\end{figure}

\begin{figure}
  \includegraphics[scale=0.6]{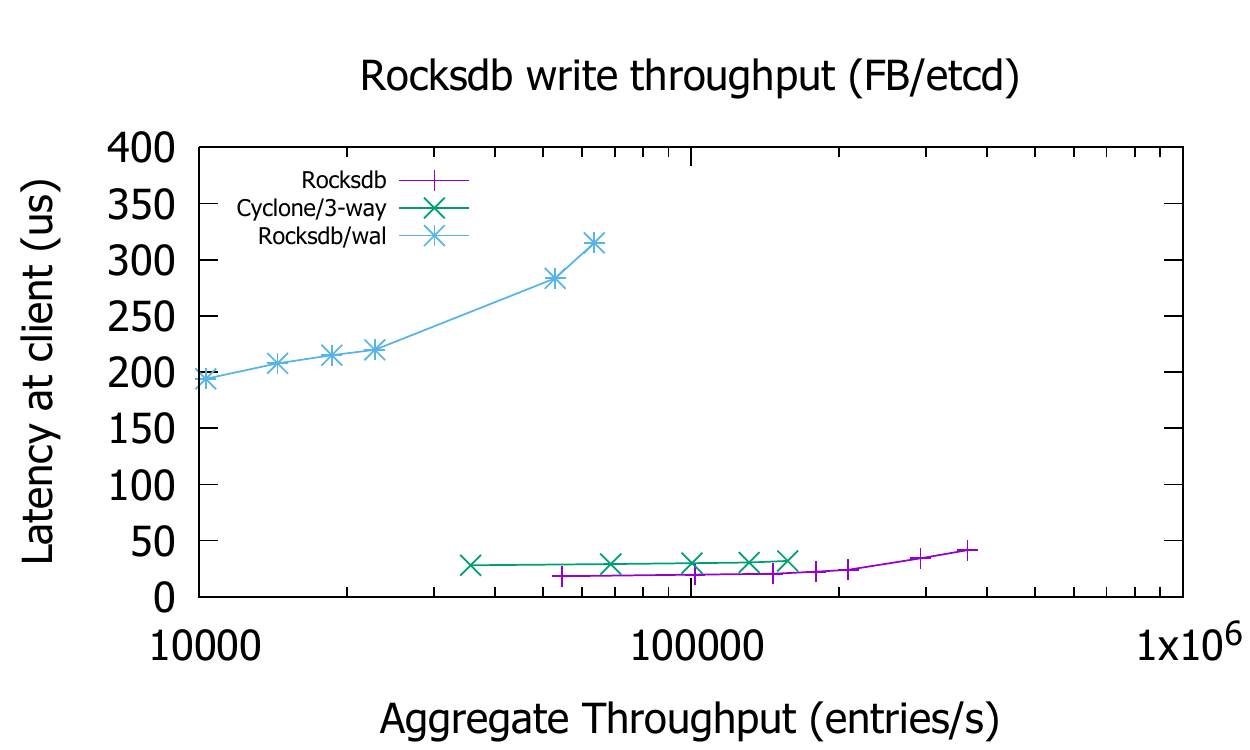}
  \caption{Rocksdb - write heavy}
  \label{fig:fb2}
\end{figure}

Our evaluation has, thus far, focused on synthetic write heavy workloads. To
illustrate that Cyclone can also be useful for real-world read-heavy workloads, 
we configured our clients to generate key value requests that mimic the
distribution of value sizes in Facebook's ETCD
trace~\cite{fb_workload_analysis,fb_scaling_memcache} and are read heavy (95\%
reads). We use weak reads in Cyclone to avoid the extra round trip for quorum reads.
Figure~\ref{fig:fb} illustrates that even with a read dominated
workload, Cyclone is effective at providing a write ahead log with 3 way
replication with a lower performance overhead than Rocksdb's own write ahead
logging. We consider the case of another real-world workload that is more write 
intensive (at 80\%) and with a smaller range of value sizes -- derived from
Facebook's VAR trace~~\cite{fb_workload_analysis,fb_scaling_memcache}, in
Figure~\ref{fig:fb2}. One can similarly conclude that Cyclone provides more
functionality (replicated write ahead log) at a far lower performance penalty
that Rocksdb's existing single machine write ahead log solution.

\begin{figure}
  \includegraphics[scale=0.6]{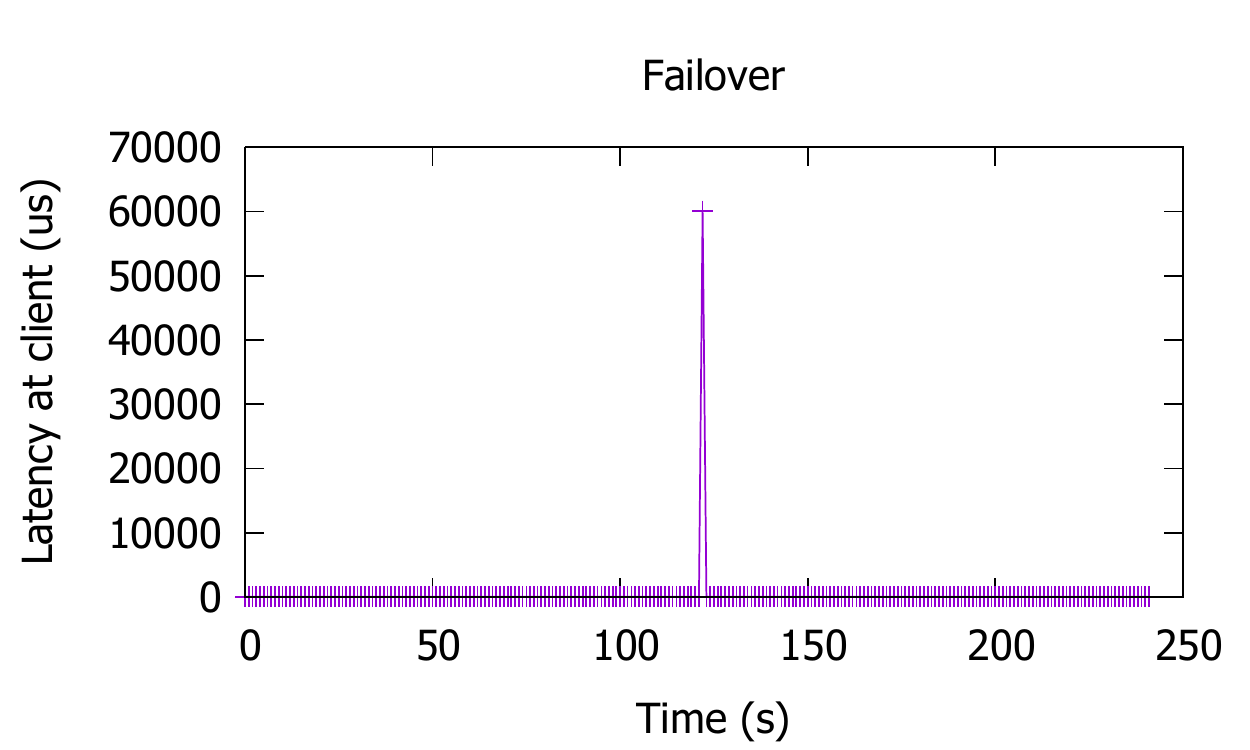}
  \caption{Rocksdb - failover}
  \label{fig:timeline}
\end{figure}

Finally, we showcase the benefit of using Cyclone beyond pure performance as
compared to the existing single machine Rocksdb write ahead log. Cyclone brings
multi-machine availability with the ability to automatically failover. We
demonstrate this in Figure ~\ref{fig:timeline} that shows the timeline of a run
where we kill the server process on the leader replica. Cyclone is configured 
with a 30ms failure detection timeout after which the client library tries
replicas in turn to locate the leader. In this case it fails over in about
60ms.

\section{Related Work}

Cyclone takes a software only approach to improving replication performance
using commodity network hardware. In contrast, Mojim~\cite{mojim} pairs directly
attached persistent memory with RDMA to replicate persistent data
structures. None of the \emph{evaluated} modes of Mojim provide any means to
keep more than two replicas in strong synchronization and therefore its
guarantees are weak compared to Cyclone.  Further, Mojim's claim that
``Paxos-like protocols'' require two networking round trips and are therefore
somehow inefficient compared to mirroring is not true. For instance, we have
demonstrated with Cyclone, where the leader responds as soon as it hears back
from a majority quorum, that the critical path to replication is just one
network round trip. We can compare Mojim's replication performance to Cyclone
for the case where we use only the NVM log in Cyclone. On 40Gbps Infiniband,
Mojim reported being able to mirror data at 4GB/s. In contrast, Cyclone can
replicate data three ways using 4*10 Gbps commodity ethernet NICs at roughly 3.7
GB/s. Note that this result is with Cyclone running the provable RAFT consensus
protocol, while Mojim is doing simple mirroring of main memory.  One can
conclude that network offload via RDMA and Infiniband is not an a-priori
requirement for high throughput replication. This is made possible by Cyclone's
approach that relieves the CPU from the burden of data copies. On the other
hand, the latency of a hop on commodity ethernet at 5 us with our DPDK stack is
larger than the close to 2us latency with Infiniband. Depending on the use case
for the key value store application the extra 6us with ethernet for the round
trip to the replica might be a concern for some. We note that mapping the
commutativity of key value store interfaces to multiple logs in Cyclone is a
technique that can also be applied to RDMA. Mojim confirms that concurrency is a
significant determinant of performance when replicating over RDMA.

Consensus in a box~\cite{consensus_box} implements the Zookeeper atomic
broadcast protocol together with a NIC on an FPGA. That work ignored durability,
focusing purely on network performance for replication. Latency is excellent due
to the fact that FPGA cuts out the path from CPU caches to the NIC. The authors
report latency as low as 3us for a round trip from leader to follower replica,
compared to the 6us we observe with DPDK. Notably however they reported 7us when
using TCP/IP from the FPGA rather than their own connection oriented network
protocol. On the other hand their peak replication throughput is 4M
replications/sec on 66 Gbps of aggregate network bandwidth for small messages,
comparable to the 6M replications/sec mark obtained by Cyclone on an aggregate
40 Gbps of network bandwidth when \emph{also persisting the log on the SSD}. We
believe the general applicability afforded by our software only approach
together with the fact that we provide persistence by design makes it a
compelling alternative to an FPGA based solution, even given the larger
latency. Also, these results demonstrate that a general purpose CPU core can be
as efficient as an FPGA in running a consensus protocol using the techniques
described in this paper.

DARE~\cite{dare} implements state machine replication over RDMA, using a custom
replication protocol. DARE does not consider persistence in its design. It is
worthwhile to point out that since DARE allows any replica to become the leader
it requires a log adjustment protocol for the leader to bring its logs up to
date, a situation we avoid by using RAFT as described in
Section~\ref{sec:dm}. Using 40 Gb/s speeds, and 3-way replication DARE reported
being able to replicate 500K requests/sec. This is without persisting the log of
updates. In contrast, Cyclone replicates upwards of 6 million similarly sized
requests per second with 4*10 Gbps ethernet (without the second level
flashlog). DARE reports an unloaded write latency of 15us, which is lower than
the unloaded write latency of Cyclone at 20us, thanks to Infiniband. The
conclusions we can draw are similar to the comparison with Mojim: a carefully
implemented log replication system obviates the need for network offload when
doing replication.

{CRANE}~\cite{crane} is a system for replicating multithreaded programs using
Paxos. It uses deterministic multi-threading to ensure replicas converge to the
same state as opposed to semantic equivalence in Cyclone using API
commutativity.  CRANE reports around a 2X slowdown for MySQL, likely rendering
it unsuitable as a replacement for the write ahead log in key value stores.

%The scalable commutativity rule~\cite{scalable_commutativity} states that for a
%shared memory program to scale in performance with concurrency it must also have
%an interface that is  for common case invocations. We have also used
%interface level commutativity to extract parallelism for the purpose of scaling
%replication performance. Others have used the same commutativity to create
%per-cpu data structures to scale KV store performance~\cite{flodb}.

Database practitioners have considered the utility of byte addressable
non-volatile memory in improving logging for
databases~\cite{nvram_log}. However, that work proposes placing the entire log
in NVM, presuming sufficient availability of such memory. It also does not
consider replication.

Cyclone does not provide exactly once semantics as we
believe that exactly once semantics are better provided by the key value store
itself by persisting necessary request markers - as done in systems such as
Kafka~\cite{kafka}. A possible direction of future work however is to add
exactly once semantics to Cyclone's client-server protocol to serve as a
building block for distributed transactions across key-value shards, each
running Cyclone, in a manner similar to other systems~\cite{raft_lin}.

\section{Conclusion}
Cyclone shows how one can leverage a small amount of non-volatile memory to
address two fundamental difficulties in adding high availability to persistent
key value stores. First, Cyclone avoids
paying the cost of the block IO interface for every update operation. This
removes block IO latency from the task of appending updates to the log as well
as sending it out over the network for replication. In turn, this enables us to
apply software packet switching techniques and concurrency to improve the
performance of replication, and making full use of available resources when
replicating log entries over the network. By optimizing the storage and network
components simultaneously, Cyclone enables high availability to persistent key
value stores without compromising on their performance.

Cyclone is available as open source software at:\\
\url{https://github.com/sdulloor/cyclone/}
\newcommand\myurl[2]{\url{#1}}
\bibliographystyle{plain}
\bibliography{paper}

\end{document}